\documentclass[aps,prl,reprint,balancelastpage,nofootinbib,preprintnumbers,superscriptaddress]{revtex4-1}

\usepackage{slashed}
\usepackage{bm}
\usepackage{latexsym,amssymb,amsmath,float,url,mathrsfs}
\usepackage{latexsym}
\usepackage{graphicx}
\usepackage{epstopdf}
\usepackage{comment}
\usepackage{subfigure}
\usepackage{natbib}
\usepackage{hyperref}
\usepackage{pifont}
\usepackage{blindtext}
\usepackage{adjustbox}
\usepackage{multirow}
\usepackage{tabularx}
\usepackage[table]{xcolor}
\usepackage{orcidlink}
\usepackage{tikz}
\usetikzlibrary{tikzmark}
\usepackage{fontawesome}
\def\draftlabel#1{{\@bsphack\if@filesw {\let\thepage\relax
			\xdef\@gtempa{\write\@auxout{\string
					\newlabel{#1}{{\@currentlabel}{\thepage}}}}}\@gtempa
		\if@nobreak \ifvmode\nobreak\fi\fi\fi\@esphack}
	\gdef\@eqnlabel{#1}}
\def\@eqnlabel{}
\def\@vacuum{}
\def\draftmarginnote#1{\marginpar{\raggedright\scriptsize\tt#1}}

\def\draft{\oddsidemargin -.5truein
		\def\@oddfoot{\sl preliminary draft \hfil
			\rm\thepage\hfil\sl\today\quad\militarytime}
		\let\@evenfoot\@oddfoot \overfullrule 3pt
		\let\label=\draftlabel
		\let\marginnote=\draftmarginnote
		\def\@eqnnum{(\theequation)\rlap{\kern\marginparsep\tt\@eqnlabel}%
			\global\let\@eqnlabel\@vacuum}  }
	
	\def\preprint{\twocolumn\sloppy\flushbottom\parindent 2em
		\leftmargini 2em\leftmarginv .5em\leftmarginvi .5em
		\oddsidemargin -.5in    \evensidemargin -.5in
		\columnsep .4in \footheight 0pt
		\textwidth 10.in        \topmargin  -.4in
		\headheight 12pt \topskip .4in
		\textheight 6.9in \footskip 0pt
		\def\@oddhead{\thepage\hfil\addtocounter{page}{1}\thepage}
		\let\@evenhead\@oddhead \def\@oddfoot{} \def\@evenfoot{} }
	
	\def\numberbysection{\@addtoreset{equation}{section}
		\def\theequation{\thesection.\arabic{equation}}}
	
	\def\underline#1{\relax\ifmmode\@@underline#1\else
		$\@@underline{\hbox{#1}}$\relax\fi}

	\def\titlepage{\@restonecolfalse\if@twocolumn\@restonecoltrue\onecolumn
		\else \newpage \fi \thispagestyle{empty}\c@page\z@
		\def\thefootnote{\fnsymbol{footnote}} }
	
	\def\endtitlepage{\if@restonecol\twocolumn \else \newpage \fi
		\def\thefootnote{\arabic{footnote}}
		\setcounter{footnote}{0}}  

\catcode`@=12
\relax

\def\figcap{\section*{Figure Captions\markboth
		{FIGURECAPTIONS}{FIGURECAPTIONS}}\list
	{Figure \arabic{enumi}:\hfill}{\settowidth\labelwidth{Figure
			999:}
		\leftmargin\labelwidth
		\advance\leftmargin\labelsep\usecounter{enumi}}}
 \relax
\def\tablecap{\section*{Table Captions\markboth
		{TABLECAPTIONS}{TABLECAPTIONS}}\list
	{Table \arabic{enumi}:\hfill}{\settowidth\labelwidth{Table
			999:}
		\leftmargin\labelwidth
		\advance\leftmargin\labelsep\usecounter{enumi}}}
 \relax
\def\reflist{\section*{References\markboth
		{REFLIST}{REFLIST}}\list
	{[\arabic{enumi}]\hfill}{\settowidth\labelwidth{[999]}
		\leftmargin\labelwidth
		\advance\leftmargin\labelsep\usecounter{enumi}}}
 \relax

	\makeatletter
	\newcounter{pubctr}
	\def\publist{\@ifnextchar[{\@publist}{\@@publist}}
	\def\@publist[#1]{\list
		{[\arabic{pubctr}]\hfill}{\settowidth\labelwidth{[999]}
			\leftmargin\labelwidth
			\advance\leftmargin\labelsep
			\@nmbrlisttrue\def\@listctr{pubctr}
			\setcounter{pubctr}{#1}\addtocounter{pubctr}{-1}}}
	\def\@@publist{\list
		{[\arabic{pubctr}]\hfill}{\settowidth\labelwidth{[999]}
			\leftmargin\labelwidth
			\advance\leftmargin\labelsep
			\@nmbrlisttrue\def\@listctr{pubctr}}}
	 \relax
	\makeatother
	\newskip\humongous \humongous=0pt plus 1000pt minus 1000pt

	\newif\ifdtup

	\relax
	
	\def\be{\begin{equation}}
		\def\ee{\end{equation}}
	\def\ba{\begin{eqnarray}}
		\def\ea{\end{eqnarray}}

\newcommand{\keff}{\kappa_{\rm eff}}
\newcommand{\Teff}{T_{\rm eff}}

	\def\IR{\relax{\rm I\kern-.18em R}}
	\def\IL{\relax{\rm I\kern-.18em L}}
	
	\def\inv{^{\raise.15ex\hbox{${\scriptscriptstyle -}$}\kern-.05em 1}}

	\def\bea{\begin{eqnarray}}
		\def\eea{\end{eqnarray}}

	\newcommand{\la}[1]{\label{#1}}

	\definecolor{markcolor2}{rgb}{1,0,0}
	
	\definecolor{markcolor3}{rgb}{0,1,0}

\usepackage{blkarray}
\usepackage{graphicx}
\usepackage{amsfonts}
\usepackage{soul}
\usepackage{amssymb}
\usepackage{amsmath}
\usepackage{cancel}
\usepackage{tcolorbox}

\usepackage{graphics,appendix,afterpage,makecell} 

\def\eq#1{{Eq.~(\ref{#1})}}

\def\fig#1{{Fig.~\ref{#1}}}

\def\dd{{\rm d}}

\definecolor{oucrimsonred}{rgb}{0.6, 0.0, 0.0}
\definecolor{persianblue}{rgb}{0.11, 0.22, 0.73}
\definecolor{forestgreen}{rgb}{0.13,0.35,0.13}
\definecolor{lightgray}{rgb}{0.83, 0.83, 0.83}
 \hypersetup{colorlinks, citecolor=oucrimsonred, linkcolor=black, urlcolor=oucrimsonred}
\definecolor{cornellred}{rgb}{0.7, 0.11, 0.11}
\definecolor{navyblue}{rgb}{0.0, 0.0, 0.5}
\definecolor{amethyst}{rgb}{0.6, 0.4, 0.8}
\definecolor{yellow}{rgb}{1.0, 1.0, 0.0}
\definecolor{firebrick}{rgb}{0.7, 0.13, 0.13}
\definecolor{tangerineyellow}{rgb}{1.0, 0.8, 0.0}
\definecolor{deepfuchsia}{rgb}{0.76, 0.33, 0.76}
\definecolor{amber}{rgb}{1.0, 0.75, 0.0}
\definecolor{VioletRed4}{rgb}{0.55, 0.13, .32}
\definecolor{indiagreen}{rgb}{0.07, 0.53, 0.03}
\definecolor{VioletRed4}{rgb}{0.55, 0.13, .32}

\definecolor{oucrimsonred}{rgb}{0.6, 0.0, 0.0}
\newcommand\vertarrowbox[3][6ex]{%
  \begin{array}[t]{@{}c@{}} #2 \\
  \left\uparrow\vcenter{\hrule height #1}\right.\kern-\nulldelimiterspace\\
  \makebox[0pt]{\scriptsize#3}
  \end{array}%
}

\definecolor{verdechiaro}{rgb}{0.6,1,0.6}
\definecolor{giallochiaro}{rgb}{1,1,0.6}
\definecolor{bluscuro}{rgb}{0.15, 0.2, 0.9}
\definecolor{verdes}{rgb}{0.1, 0.5, 0.1}%
\definecolor{tangerineyellow}{rgb}{1.0, 0.8, 0.0}

\definecolor{americanrose}{rgb}{1.0, 0.01, 0.24}
\definecolor{cobalt}{rgb}{0.0, 0.28, 0.67}
\definecolor{brandeisblue}{rgb}{0.0, 0.44, 1.0}
\definecolor{mycolor}{rgb}{0.0, 0.0, 0.5}
\definecolor{oxfordblue}{rgb}{0.0, 0.13, 0.28}
\definecolor{azure}{rgb}{0.0, 0.5, 1.0}
\definecolor{turquoiseblue}{rgb}{0.0, 1.0, 0.94}
\newtcolorbox{mynewbox}[1]{colback=white!5!white,colframe=azure!75!black,fonttitle=\bfseries,title=#1}
\newtcolorbox{mybox}{colback=mycolor!5!white,colframe=azure!75!black}
\newtcolorbox{mynamedbox}[1]{colback=mycolor!5!white,colframe=azure!75!black,title=#1}
\definecolor{venetianred}{rgb}{0.78, 0.03, 0.08}
\newtcolorbox{mynamedbox1}[1]{colback=venetianred!5!white,colframe=venetianred!80!black,title=#1}
\newtcolorbox{mynamedbox2}[1]{colback=azure!5!white,colframe=azure!80!black,title=#1}

\definecolor{verdes}{rgb}{0.1, 0.5, 0.1}%
\definecolor{cornellred}{rgb}{0.7, 0.11, 0.11}

\definecolor{VioletRed4}{rgb}{0.55, 0.13, .32}

\hypersetup{
     colorlinks   = true,
     citecolor    = violet,
     urlcolor     = violet,
     linkcolor    = violet}

\definecolor{rossocorsa}{rgb}{0.83, 0.0, 0.0}

\usepackage[normalem]{ulem}

\begin{document}

\title[]{Black Hole Mergers as the Fastest Photon Ring Scramblers
}

\author{D.~Giataganas\orcidlink{0000-0003-2003-3902}
}
\affiliation{Department of Physics, National Sun Yat-Sen University, Kaohsiung 80424, Taiwan}
\affiliation{Physics Division, National Center for Theoretical Sciences, Taipei 10617, Taiwan}

\author{G.F.~Giudice\orcidlink{0000-0002-0247-4096}}
\affiliation{New York University, Abu Dhabi, PO Box 129188 Saadiyat Island, Abu Dhabi, UAE}
\affiliation{CERN, Theoretical Physics Department, Geneva, Switzerland}

\author{A.~Ianniccari\orcidlink{0009-0008-9885-7737}}
\affiliation{Department of Theoretical Physics and Gravitational Wave Science Center,  \\
24 quai E. Ansermet, CH-1211 Geneva 4, Switzerland}

\author{A.J.~Iovino\orcidlink{0000-0002-8531-5962}}
\affiliation{New York University, Abu Dhabi, PO Box 129188 Saadiyat Island, Abu Dhabi, UAE}

\author{A.~Kehagias\orcidlink{0000-0001-6080-6215}}
\affiliation{Physics Division, National Technical University of Athens, Athens, 15780, Greece}

\author{F.~Quevedo\orcidlink{0000-0002-7810-3662}}
\affiliation{New York University, Abu Dhabi, PO Box 129188 Saadiyat Island, Abu Dhabi, UAE}
\affiliation{DAMTP, University of Cambridge, Wilberforce Road, Cambridge CB2-0WA, UK}

\author{D.~Perrone\orcidlink{0000-0003-4430-4914}}
\affiliation{Department of Theoretical Physics and Gravitational Wave Science Center,  \\
24 quai E. Ansermet, CH-1211 Geneva 4, Switzerland}

\author{A.~Riotto\orcidlink{0000-0001-6948-0856}}
\affiliation{Department of Theoretical Physics and Gravitational Wave Science Center,  \\
24 quai E. Ansermet, CH-1211 Geneva 4, Switzerland}
\affiliation{INFN, Sezione di Roma, Piazzale Aldo Moro 2, 00185, Roma, Italy}


\begin{abstract}
\noindent
Black holes are the most efficient scramblers in nature. By mapping the instantaneous mass and angular momentum of two spinless black holes in a quasi-circular binary onto those of an effective Kerr black hole, we demonstrate that the final state of the merger remnant corresponds with remarkable accuracy to the configuration that renders  null geodesics unstable at the highest possible rate. This suggests a deep connection between the properties of black holes resulting from binary mergers and their unstable null orbits. 
\end{abstract}

\maketitle

\section{Introduction}
The extreme nature of black holes is revealed by their tendency to saturate theoretical bounds on various physical quantities. For instance, black holes contain the maximum possible entropy that can be stored in a given region of space, thus providing the ultimate repository of information in the universe, as exhibited by the Bekenstein bound~\cite{Bekenstein:1980jp} and the covariant entropy bound~\cite{Bousso:1999xy}. They are characterized by extremely strong scattering, as reflected by the viscosity bound~\cite{Kovtun:2004de}.
Moreover, it has been conjectured that black holes are the fastest possible scramblers of information. This is because they thermalize information in a time that grows only logarithmically with the number of degrees of freedom in the system. This property singles out black holes as the most efficient scramblers in nature~\cite{Hayden:2007cs,Sekino:2008he} and as the physical objects that saturate the quantum chaos bound \cite{Maldacena:2015waa}. 

A particularly interesting testing ground for the extreme properties of black holes comes from massive binary mergers, which is a phenomenon accessible to experimental observations through gravitational waves. Recently, the authors of Ref.~\cite{Rincon-Ramirez:2026tbo} have conjectured that the final state of binary black hole mergers is uniquely determined by the condition of maximal entropy. Once the instantaneous total mass (including binding energy) and angular momentum of quasi-circular, non-spinning binaries are mapped onto those of an equivalent Kerr black hole, the corresponding entropy attains a maximum at a specific stage of the evolution. Strikingly, this maximum occurs at values of mass and angular momentum very close to those predicted by numerical relativity for the merger remnant. This observation suggests that a simple thermodynamic principle, the maximization of the entropy of a system subject to balance laws for the energy and angular momentum, selects the properties of the final remnant.

In this letter we explore another remarkable property of black hole dynamics and we conjecture that the final state in binary mergers achieves the shortest  scrambling time for the photon shell surrounding the black hole to spread information. We demonstrate our claim by showing that the mass and spin of the Kerr black hole formed from the merger of two non-spinning black holes can be determined with high accuracy by maximizing the averaged Lyapunov coefficient of the remnant black hole photon shell, which characterizes the rate at which the system spreads information away from the photon shell, using Mino time. This result reveals a compelling correspondence between fundamental black hole properties (mass and spin), their null geodesics (which characterize the photon shell), and the nonlinear gravitational dynamics that give rise to the final state.

Geodesics are known to carry important information about the black hole structure. For instance, the binding energy associated with the innermost stable circular time-like orbit in Kerr spacetime provides a powerful method to infer the spin of astrophysical black holes~\cite{Zhang:1997dy,Narayan:2005ie,Shapiro}. The unstable null geodesics that characterize the photon ring determine the observational appearance of black holes as seen by distant observers and have been linked to the characteristic quasi-normal mode spectra~\cite{Press,Nollert_1999,Kokkotas:1999bd,Berti:2025hly}. These modes admit an interpretation in terms of massless particles temporarily trapped near the unstable circular null orbit before gradually leaking away~\cite{Goebel,Ferrari:1984zz,Mashhoon:1985cya,Berti:2005eb}. The real part of the quasi-normal frequency is determined by the angular velocity evaluated at the unstable null geodesic, whereas the imaginary part has been shown to correlate with the instability timescale of that orbit~\cite{Cornish:2003ig,Cardoso:2008bp}. This characteristic timescale is governed by the Lyapunov exponent, which measures the exponential divergence of nearby trajectories. Crucially, the photon ring Lyapunov exponent is not merely a diagnostic of classical instability: as shown in Refs.~\cite{Giataganas:2024hil,Giataganas:2026ctn}, it is the unique quantity that saturates the quantum chaos bound~\cite{Maldacena:2015waa}, $\lambda = 2\pi k_B T_{\rm ind}/\hbar$, where $T_{\rm ind}$ is the Unruh temperature of the Rindler geometry induced on a probe string worldsheet in the near-ring region. This property is specific to the photon ring and has no analogue for the Bekenstein--Hawking entropy, analyzed in Ref.\,\cite{Rincon-Ramirez:2026tbo}, which is associated with the event horizon and does not saturate the quantum chaos bound in the same geometric sense.

Unstable circular orbits may also shed light on phenomena occurring near the threshold of black hole formation in high-energy scattering processes, and additional indications point to a relation between the scaling exponent controlling the number of revolutions completed by two Schwarzschild black holes prior to merging into a Kerr black hole and the Lyapunov exponent of circular geodesics in the resulting Kerr spacetime~\cite{Pretorius:2007jn}. Moreover, the critical threshold of the compaction function required to form a black hole during the radiation epoch in the early universe is well approximated by the threshold for the emergence of the first unstable circular orbit in a spherically symmetric background. Additionally, the critical exponent governing the scaling law of black hole masses close to threshold is determined by the inverse of the Lyapunov coefficient of the unstable orbits, when a self-similar phase develops near criticality~\cite{Ianniccari:2024ltb}. 

Taken together, all these considerations suggest that the properties of null geodesics in the final black hole configuration may be closely intertwined with the nonlinear dynamics governing its formation. The results presented in this letter corroborate this conclusion.

Throughout the paper, we set $G_N=c=1$.

\vskip 0.3cm
\section{PN expansion, Kerr black hole, and the Lyapunov coefficients of its photon shell}
Within the post-Newtonian (PN) framework, the physical characteristics of a black hole binary are expressed as a power series in a small expansion parameter. A convenient choice is the dimensionless quantity $x=(M_\mathrm{tot}\Omega)^{2/3}$,
where 
$\Omega$ denotes the orbital frequency and $M_\mathrm{tot}=M_1+M_2$ is the sum of the individual masses $M_1$ and $M_2$ of the two non-spinning black holes.
Quantities such as the orbital angular momentum $J(x)$, the binding energy $E(x)$, and the total mass of the system $M(x)=M_\mathrm{tot}+E(x)$ can be expanded in powers of $x$, with $0 < x \ll 1$. As gravitational radiation is emitted, both $M$ and $J$ decrease monotonically with time, while the orbital frequency $\Omega$ increases; within this approximation, $M_1$ and $M_2$ remain constant.
Therefore, one may eliminate the parameter $x$ from the parametric curve $[M(x), J(x)]$ and express the evolution of the mass directly as a function of the angular momentum, treating $J$ as an effective time variable. Working consistently to fourth post-Newtonian order (4PN), one obtains the relation $M(j)$ for quasi-circular orbits \cite{Damour:2014jta} (see also Ref.~\cite{Damour:2016bks} for a review)
\begin{eqnarray}
M(j)\hspace{-0.2cm}&=&\hspace{-0.2cm}M_\mathrm{tot}\hspace{-0.1cm}
-\hspace{-0.1cm}\frac{\nu M_\mathrm{tot}}{2j^2}
\left\{1+\left( \frac{9+\nu}{4} \right)\frac{1}{j^2}
+\frac{81-7\nu+\nu^2}{8}\frac{1}{j^4}\right. \nonumber\\
&+&\left[\frac{3861}{64}
+\left(\frac{41\pi^2}{32}-\frac{8833}{192}\right)\nu
-\frac{5\nu^2}{32}
+\frac{5\nu^3}{64}\right]\frac{1}{j^6} \nonumber\\
&+&\left[\left(\frac{6581\pi^2}{512}
-\frac{989911}{1920}
-\frac{128}{5}\left(\gamma_E+\ln\frac{4}{j}\right)\right)\nu \right. \nonumber\\
&+&\left(\frac{8875}{384}
-\frac{41\pi^2}{64}\right)\nu^2
-\frac{3\nu^3}{64}
+\frac{7\nu^4}{128}
+\frac{53703}{128}\Bigg]\frac{1}{j^8}
\Bigg\}, \nonumber\\
&&
\label{eq:Mj}
\end{eqnarray}
where $\nu = M_1 M_2 / M_\mathrm{tot}^2$ denotes the symmetric mass ratio. We have also introduced the standard dimensionless angular momentum parameter $j$, defined through
\be
J(j) = M_\mathrm{tot}^2 \nu j .
\ee
This formulation eliminates the intermediate PN parameter $x$ and provides a direct representation of the system’s mass as a function of the dimensionless angular momentum $j$, which can be regarded as an alternative evolutionary parameter.

Following the prescription of Ref.~\cite{Rincon-Ramirez:2026tbo}, at any given instant we map the binary system onto a Kerr black hole with mass $M(j)$ and spin $J(j)$.
For a generic Kerr black hole, the photon shell is defined as the region of spacetime containing unstable null geodesics confined to bound trajectories that neither escape to infinity nor cross the event horizon for extended periods.
In the Schwarzschild limit, this structure reduces to a two-dimensional spherical surface at $r = 3M$. In contrast, for a Kerr black hole, the photon region extends into a three-dimensional spherical shell. A convenient description employs Boyer–Lindquist coordinates (see App. A), in which all bound photon trajectories remain at constant radius within the interval $r_-\leq r\leq r_+$,
where
\begin{equation}
r_\pm = 2M\left[1+\cos\left(\frac{2}{3}\arccos\left(\pm\frac{a}{M}\right)\right)\right],
\end{equation}
and the spin parameter is $a = J/M$.
Within the equatorial annulus defined by $r_- \le r \le r_+$ and $\theta = \pi/2$, each point corresponds to exactly one bound null orbit. At the limiting radii $r = r_\pm$, motion is completely restricted to the equatorial plane. However, for intermediate radii, the trajectories display oscillatory motion in the polar direction, varying between the turning angles
\begin{align}
\label{eq::Turning_Points}
\theta_\pm = \arccos\left(\mp\sqrt{u_+}\right),
\end{align}
with
\begin{eqnarray}
u_\pm(r) &=& \frac{r}{a^2(r-M)^2}
\left[
-r^3+3M^2r-2a^2M\right.
\nonumber\\
&\pm&\left.
2\sqrt{M\Delta(2r^3-3Mr^2+a^2M)}
\right],
\nonumber\\
\Delta&=& r^2-2Mr+a^2.
\end{eqnarray}
We refer to a complete oscillation in $\theta$, for example from $\theta_-$ back to $\theta_-$, as a single orbit. Since the azimuthal coordinate $\phi$ advances during each cycle, the photon generally returns close to, though not precisely at, its initial spatial location.
Thus, the photon shell occupies the spacetime region
\begin{align}
r_- \le r \le r_+,\quad \theta_- \le \theta \le \theta_+,\quad 0 \le \phi < 2\pi,
\end{align}
for all times $-\infty < t < \infty$.
These bound null geodesics are unstable: even an arbitrarily small perturbation causes the photon either to plunge into the black hole or to escape outward toward distant observers. The observed photon ring arises from light rays that follow such nearly bound trajectories, but ultimately escape.
To characterize this instability, consider two neighboring geodesics, one exactly bound and one initially separated by an infinitesimal radial displacement $\delta r_0$. The geodesic deviation equation implies that after $n$ half-oscillations between $\theta_\pm$, their radial separation becomes
\begin{align}
\label{eq::Geodesic_Deviation}
\delta r_n = e^{\gamma n}\delta r_0.
\end{align}
The parameter $\gamma$ is the dimensionless per-half-orbit instability exponent, which depends on the specific bound orbit and is given by \cite{Kapec:2019hro, Johnson:2019ljv}
\be
\label{eq::Lyapunov_Exponent}
\gamma(r) = \sqrt{\frac{{\cal R}''(r)}{2}} \cdot G_\theta(r), ~~~~
G_\theta(r) = \int_{\theta_{-}}^{\theta_+}
\frac{{\rm d}\theta}{\sqrt{\Theta(\theta)}},
\ee
where ${\cal R}(r)$ and $\Theta(\theta)$ denote the radial and polar potentials, respectively, both provided in App. C, and primes indicate differentiation with respect to the radius. The function $G_\theta$ corresponds
to the so-called ``Mino time"  $\tau_{\rm M}$ \cite{Mino:2003yg}, which is the time interval required to complete a half-orbit for a null
geodesic around a rotating black hole. It is the natural clock along the geodesics, as it ticks uniformly along the paths, allowing the radial and polar equations to decouple. Therefore, the  instability parameter can be expressed as a function of the Lyapunov coefficient in Mino time (defined in App. C), 
\be
\gamma(r)=\lambda_{\rm p}(r)\cdot \tau_{\rm M}(r),\,\,\, \lambda_{\rm p}(r)=
\sqrt{\frac{{\cal R}''(r)}{2}}.
\ee
These Lyapunov exponents, one for each radius $r$, encode the instability properties of the Kerr photon shell and  measure the rate of instability  production.
For the photon shell of a Kerr black hole, we integrate over all unstable trajectories and average over the Mino time necessary to perform the allowed half-orbits   
\be
\overline{\lambda}_{\rm p}\left[M(j),J(j)\right]=\frac{\int_{r_-}^{r_+}{\rm d}r \,\gamma(r)}{\int_{r_-}^{r_+}{\rm d}r \,G_\theta(r)}=\frac{\int_{r_-}^{r_+}{\rm d}r \,\lambda_{\rm p}(r)\cdot \tau_{\rm M}(r)}{\int_{r_-}^{r_+}{\rm d}r \,\tau_{\rm M}(r)}.
\ee
Here we have made explicit the dependence on the mass and spin of the corresponding Kerr black hole. The parameter $\overline{\lambda}_{\rm p}$ is the averaged Lyapunov coefficient in terms of Mino time, which has the physical meaning of the typical rate to spread information away from the photon shell. Indeed, note that $\tau_{\rm M}$, in spite of its name, has the dimension of an inverse time.

It is  worth mentioning a possible  connection between $\overline{\lambda}_{\rm p}$ and the Kolmogorov-Sinai entropy \cite{Kolmogorov1959} which, for an unstable system, is expressed as the sum of the Lyapunov coefficients through Pesin's entropy formula~\cite{pesin1977characteristic} 
\be
h_{\rm KS}=\sum_{\lambda_i>0}\lambda_i. 
\ee
The entropy $h_{\rm KS}$ measures the rate at which information is generated, which has a close correspondence to $\overline{\lambda}_{\rm p}$ when Mino time is considered. This might be related to the fact that, taking the Penrose limit 
along the photon shell of a Kerr black hole, the corresponding Hamiltonian contains an unstable  quadratic potential   \cite{Hadar:2022xag,Fransen:2023eqj,Giataganas:2024hil,Perrone:2025zhy} and the Kolmogorov-Sinai entropy should grow linearly in time \cite{Bianchi:2017kgb}.
A deep understanding of this further connection is left for future work.
\vskip 0.3cm
\section{The final state from the null geodesics}
\begin{figure*}[t!]
\centering
  \includegraphics[width=0.99\textwidth]{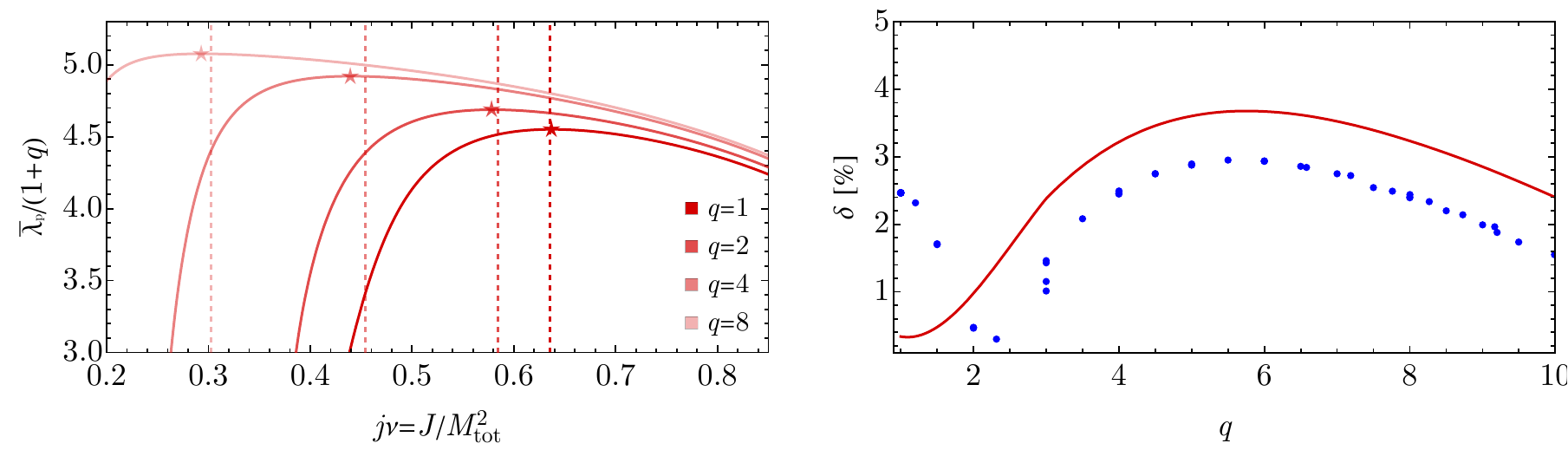}
  \caption{\textit{Left Panel:} The averaged
Lyapunov coefficient $\overline{\lambda}_{\rm p}$, rescaled by  $(1+q)$, as a function of the binary angular momentum for several mass ratios $q$. Stars indicate the locations of the maxima. The vertical lines indicate the values obtained from the numerical fit in Eq. (\ref{fit}), once the final binary mass has been rescaled from  $M_f=M(j_*)$ to $M_\mathrm{tot}$.  \textit{Right Panel:} Relative difference $\delta$ between the final angular momentum obtained from the NR fitting formula of Eq.~(\ref{fit}) (red solid line) and the SXS catalog data (blue dots), and the value $\nu j_*$ that maximizes $\overline{\lambda}_{\rm p}$, as a function of the mass ratio $q$. The SXS data points are drawn from the subset of 60~simulations satisfying the selection criteria described in the text; the full table is provided in App.~E.}
  \label{fig:1}
\end{figure*}

Our conjecture affirms that the final-state configuration in black hole binary mergers is determined by the condition of fastest information spreading away from the photon shell. Quantitatively, this translates into the condition of maximizing $\overline{\lambda}_{\rm p}$. The result of this procedure is shown in the left panel of \fig{fig:1}, where we plot $\overline{\lambda}_{\rm p}$ as a function of the binary angular momentum $j\nu$, for different values of $q\equiv M_2/M_1$ (such that $\nu= q/(1+q)^2$), and identify the location of the maxima with final spin $J(j_*)/M_{\rm tot}^2=\nu j_*$.
\noindent
\hspace{-0.2cm}
Next, we compare our prediction with numerical simulations for non-spinning binaries in quasi-circular motion, which accurately describe the final spin in terms of a polynomial in $\nu$ \cite{Hofmann:2016yih}
\begin{eqnarray}
\label{fit}
\frac{a_f}{M_f}&=&\frac{J_f}{M_f^2}
\simeq 2\sqrt{3}\nu - 3.87 \nu^2 +3.39 \nu^3\nonumber\\
&+&4.49\nu^4-5.77\nu^5 -13.04 \nu^6.
\end{eqnarray}
Here, $M_f$ and $J_f$ denote the final mass and spin of the black hole after merger.
The vertical dashed lines in \fig{fig:1} indicate the values obtained by the numerical fit in \eq{fit}, once the total mass $M_f=M(j_*)$ is properly rescaled to $M_\mathrm{tot}$. In the right panel of \fig{fig:1}, we show the relative difference
\begin{equation}
\delta\equiv \frac{J_{f}/M_\mathrm{tot}^2-\nu j_*}{\nu j_*}
\end{equation}
between the values of the black hole spins derived from numerical fits of Eq.\,\eqref{fit} (red lines) and from maximizing the averaged Lyapunov coefficient. In addition, we compare our predictions against direct NR results from the SXS Collaboration's third binary black hole catalog~\cite{Scheel:2025jct} (blue dots), selecting the subset of simulations corresponding to non-spinning, quasi-circular inspirals (see App.~E for the full selection criteria and data).\footnote{A complementary comparison using surrogate waveform models~\cite{Varma:2018mmi,Yoo:2022erv} would constitute a further check of our conjecture and is left for future work.} The agreement between our prediction and the numerical fits is striking, as $\delta$ is always within a few percent for $q\lesssim 20$.

For larger values of $q$ our prescription cannot be applied at the same level of accuracy. Analytically, we find 
\begin{eqnarray}
    \overline{\lambda}_{\rm p}(q\gg 20)&\simeq& 3\sqrt{3} M(j_*)\left(1-\frac{13}{81}\,\nu^2j_*^2\; \right),\nonumber\\
\nu j_*(q\gg 20)&\simeq& \frac{3}{(26q^3)^{1/4}}, \quad \nu (q\gg 1)\simeq \frac{1}{q}.
\label{scalingq}
\end{eqnarray}
When $q$ is large, the lighter black hole acts as a test particle, leaving the heavier black hole nearly unaffected, and the probe-Schwarzschild limit is reached very quickly.
Therefore, for very large $q$, our analysis predicts that the final spin should scale like $1/q^{3/4}$ rather than $1/q$, as derived by the numerical fit in \eq{fit}.  One can also check that the same scaling for $\nu j_*$ as in \eq{scalingq} is obtained by extremizing the Kerr black hole entropy as suggested in Ref.\,\cite{Rincon-Ramirez:2026tbo}. 
 
The remaining few-percent mismatch for $q\lesssim 20$ is not surprising and can be attributed to the various uncertainties involved in our calculation. As shown in App. D, it is not implausible that the unknown fifth-order PN corrections could alter our prediction by an amount as large as a few percent. 
Moreover, using alternative numerical fits ({\it e.g.}, see Refs. \cite{Jimenez-Forteza:2016oae,Healy:2016lce}), we find variations just below the percent range. 

More importantly, our procedure has intrinsic systematic uncertainties. First, while our criterion of fast information spread has a well-defined physical meaning, the choice of averaging the Lyapunov coefficients in terms of $\overline{\lambda}_{\rm p}$ is not unique. 
Second, the procedure of mapping the binary initial state into the Kerr final state does not necessarily capture the full nonlinearity of the system's evolution and the gravitational wave emission. 

Taking all these considerations into account, we can safely claim that our conjecture is numerically compatible with state-of-the-art simulations for $q\lesssim 20$.
Therefore, we conclude that the final state of the black hole binary merger maximizes the Lyapunov coefficient in Mino time, thus minimizing the averaged  time in which the instability of the photon shell grows. 
\vskip 0.3cm
\section{Conclusions and final remarks} 
In many approaches to quantum gravity and high-energy physics, black holes appear to realize the limiting values of fundamental inequalities governing physical observables, placing them as extremal objects in theory space. In this letter, we have uncovered a new intriguing property of black holes, related to the unstable photon shell of the final state in the merging between two non-spinning black holes in quasi-circular orbit. Building on the results of Ref.~\cite{Rincon-Ramirez:2026tbo}, we have mapped the total mass and angular momentum of the binary system onto an effective Kerr geometry. Then, we have shown that the mass and spin of the physical Kerr black hole formed after the merger are determined by the requirement that the averaged instability rate of the final state photon shell be maximized. In other words, the merger selects the configuration that spreads information at the fastest possible rate. This provides yet another hint towards the fact that the null geodesics of the final state black hole somehow  encode information about the nonlinear dynamics leading to it. Our conjecture is physically distinct from the entropy-maximization proposal of Ref.~\cite{Rincon-Ramirez:2026tbo}. While entropy concerns the static storage of information behind the event horizon, $\overline{\lambda}_{\rm p}$ characterizes the dynamical scrambling of information away from the photon ring. The deeper justification for the Lyapunov criterion lies in the fact that, as demonstrated in Ref.~\cite{Giataganas:2026ctn}, the photon ring Lyapunov exponent is the unique quantity that saturates the quantum chaos bound~\cite{Maldacena:2015waa}. This saturation is a property of the photon ring alone and has no counterpart for the Bekenstein--Hawking entropy. Maximizing $\overline{\lambda}_{\rm p}$ is therefore equivalent to selecting the merger remnant that saturates the quantum chaos bound as tightly as possible throughout the photon shell, a statement without analogue for the entropy.

It will also be interesting to explore the connection with the  universal bound for the maximum average rate of information emission achievable by a physical system obtained in Ref.\,\cite{Bekenstein:1981zz} and recently tested through  observational data\,\cite{Carullo:2021yxh}.


\vspace{-0.6cm}
\begin{acknowledgments}
\section{Acknowledgments}
\vspace{-0.2cm}
We thank the authors of Ref. \cite{Rincon-Ramirez:2026tbo} for comments on the manuscript. D.G.  acknowledges support from  the National Science and Technology Council (NSTC) of Taiwan with the Young Scholar Columbus Fellowship grant 114-2636-M-110-004. A.J.I. and F.Q. research is funded by Tamkeen under the research grant to NYUAD ADHPG-AD457. A.I. and A.R.  acknowledge support from the Swiss National Science Foundation (project number CRSII5\_213497). A.R. thanks NYUAD for the kind hospitality where this project was initiated.
\end{acknowledgments}

\bibliography{main}
\clearpage
\appendix
\begin{center}

{ \it \Large Appendices}
\end{center}
\setcounter{equation}{0}
\setcounter{figure}{0}
\setcounter{section}{0}
\setcounter{table}{0}
\renewcommand{\theequation}{A\arabic{equation}}
\renewcommand{\thefigure}{A\arabic{figure}}
\renewcommand{\thetable}{A\arabic{table}}
\vspace{-0.5cm}
\section{A. Kerr metric and equatorial reduction}\label{App:A}
\noindent 
In Boyer--Lindquist coordinates $(t,r,\theta,\phi)$, the metric of  Kerr black hole with mass $M$ and spin parameter $a$ reads
\begin{align}\nonumber
{\rm d}s^2 &=
-\left(1-\frac{2Mr}{\Sigma}\right)\,{\rm d}t^2
-\frac{4Mar\sin^2\theta}{\Sigma}\,{\rm d}t\,{\rm d}\phi+\frac{\Sigma}{\Delta}\,{\rm d}r^2\\\nonumber
&+\Sigma\,{\rm d}\theta^2
+\left(r^2+a^2+\frac{2Ma^2r\sin^2\theta}{\Sigma}\right)\sin^2\theta\,{\rm d}\phi^2,
\\
\Sigma&\equiv r^2+a^2\cos^2\theta,\qquad
\Delta\equiv r^2-2Mr+a^2.
\end{align}
On the equatorial plane, $\theta=\pi/2$ we have $\Sigma=r^2$ and the nonzero components of the metric simplify as
\begin{align}\nonumber
&g_{tt}= -\left(1-\frac{2M}{r}\right),\quad
g_{t\phi}= -\frac{2Ma}{r},\\
&g_{\phi\phi}= r^2+a^2+\frac{2Ma^2}{r},\quad
g_{rr}=\frac{r^2}{\Delta}.
\label{eq:equatorial-components}
\end{align}
Kerr admits two commuting Killing fields
\begin{equation}
\xi=\partial_t,\qquad \psi=\partial_\phi.
\end{equation}
For a circular orbit with constant angular velocity $\omega={\rm d}\phi/\dd t$, we can define the helical Killing field
\begin{equation}
\chi \equiv \xi+\omega\,\psi=\partial_t+\omega\,\partial_\phi,
\label{eq:helical-killing}
\end{equation}
with a squared norm equal to 
\begin{equation}
\chi^2(r;\omega)=g_{tt}+2\omega g_{t\phi}+\omega^2 g_{\phi\phi}.
\label{eq:chi2-def}
\end{equation}
Using Eq. \eqref{eq:equatorial-components} we get the explicit equatorial expression
\begin{equation}
\chi^2(r;\omega)=
-\left(1-\frac{2M}{r}\right)
-\frac{4Ma}{r}\omega
+\left(r^2+a^2+\frac{2Ma^2}{r}\right)\omega^2.
\label{eq:chi2-equatorial}
\end{equation}
We define a co-rotating angle $\varphi$ by
\begin{equation}
{\rm d}\varphi \equiv {\rm d}\phi-\omega\,{\rm d}t.
\end{equation}
The metric restricted to ${\rm d}\varphi=0$ when $\phi=\omega t$
becomes
\begin{equation}
{\rm d}s^2\Big|_{{\rm d}\varphi=0}=\chi^2(r;\omega)\,{\rm d}t^2+g_{rr}(r)\,{\rm d}r^2.
\label{eq:2dmetric-tr}
\end{equation}
This is a Kerr analog of the Schwarzschild construction, where $\chi^2$ plays the role of the redshift factor for the helical time.
The equatorial circular null orbit at radius $r=r_{0}$ and angular velocity $\omega=\omega_{0}$ is characterized by two conditions.
First, the helical Killing field becomes null,
\begin{equation}
\chi^2(r_{0};\omega_{0})=0.
\label{eq:null-cond}
\end{equation}
Second, the circular orbit condition further requires an extremum,
\begin{equation}
\partial_r \chi^2(r_{0};\omega_{0})=0,
\label{eq:extremum-cond}
\end{equation}
or equivalently $\partial_\rho\chi^2|_{r_{0}}=0$ in proper radial distance $\rho$.
This leads to a local null Killing horizon in the reduced effective two-dimensional metric defined by Eq. \eqref{eq:null-cond}, while the extremum condition Eq. \eqref{eq:extremum-cond} implies that the linear term in the near-orbit expansion vanishes. As a result, this  metric takes a local Rindler-like form. This is equivalent to the double root structure of the effective potential for a circular null orbit.

From  Eqs. \eqref{eq:null-cond} and \eqref{eq:extremum-cond}, while restricted on the equator, we obtain
\begin{equation}
r_{0}^2-3Mr_{0}\pm 2a\sqrt{Mr_{0}}=0,
\label{eq:rph-equation}
\end{equation}
and in closed form the solutions read
\begin{equation}
r_\pm = 2M\left[1+\cos\left(\frac{2}{3}\arccos\left(\pm\frac{a}{M}\right)\right)\right].
\end{equation}
Note that, for $a\to 0$, we find $r_\pm\to 3M$ as expected, while at extremality $a=M$ we find $r_{-}=M$ for the prograde, and $r_{+}=4M$ for the retrograde. The extremum condition (\ref{eq:extremum-cond}), together with the photon ring radius equations, leads to 
\be
\omega_{0}{}_{\pm}=\frac{\sqrt{M}}{a\sqrt{M}\mp (r_{\pm})^{3/2}},
\label{eq:Omega}
\ee
and to the corresponding Lyapunov coefficients
\begin{equation}\nonumber
    \lambda_{\pm}=
\frac{\sqrt{3\,\Delta_{\pm}}}{r_{\pm}\,\sqrt{a^2+3r^2_{\pm}}}, \qquad
\Delta_\pm=r_{\pm}^2-2Mr_{\pm}+a^2.
\end{equation}
\setcounter{equation}{0}
\setcounter{figure}{0}
\setcounter{section}{0}
\setcounter{table}{0}
\renewcommand{\theequation}{B\arabic{equation}}
\renewcommand{\thefigure}{B\arabic{figure}}
\renewcommand{\thetable}{B\arabic{table}}
\vspace{-0.5cm}
\section{B. Rindler metric form and the effective temperature}
We define the proper radial distance $\rho$ at fixed $(t,\phi)$ as
\begin{equation}
{\rm d}\rho =\sqrt{g_{rr}}\,{\rm d}r=\frac{r}{\sqrt{\Delta}}\,{\rm d}r,\quad \partial_\rho = \frac{1}{\sqrt{g_{rr}}}\partial_r=\frac{\sqrt{\Delta}}{r}\partial_r.
\label{eq:rho-def}
\end{equation}
Near the photon ring radius $r=r_{0}$, using Eqs. \eqref{eq:null-cond} and \eqref{eq:extremum-cond}, we have the Rindler form of \eqref{eq:2dmetric-tr} with 
\begin{equation}
\chi^2(\rho)\;=\;\frac{1}{2}\left(\partial_\rho^2\chi^2\right)_{ r_0}\rho^2+{\cal O}(\rho^3).
\end{equation}
For an unstable photon ring, the coefficient  corresponding to the curvature of the Killing norm near the null surface has the sign appropriate to a Rindler wedge. Therefore, we define
\begin{equation}
\keff^2\equiv \frac{1}{2}\left(\partial_\rho^2\chi^2\right)\bigg|_{r_0} = \frac12\left(\frac{\Delta}{r^2}\partial_r^2\chi^2\right)\bigg|_{r_0} 
\label{eq:kappa-def}
\end{equation}
and for the unstable photon ring we get
\begin{equation}
{\rm d}s^2_{(2)} \approx -\keff^2\rho^2\,{\rm d}t^2+{\rm d}\rho^2,
\label{eq:rindler}
\end{equation}
where  $\keff$ is the effective surface gravity or acceleration parameter. This determines the associated Unruh temperature as
\begin{equation}
\Teff \equiv \frac{\keff}{2\pi}.
\label{eq:Teff-def}
\end{equation}
Using Eqs. (\ref{eq:null-cond}) and (\ref{eq:extremum-cond}), we obtain
\begin{equation}\la{eq:kappa-compact-01}
\keff
=
\frac{\sqrt{\Delta_{0}\left(1-a^2\omega_{0}^2\right)}}{r_{0}^2}
\end{equation}
with  $\Delta_{0}\equiv r_{0}^2-2Mr_{0}+a^2$, which is non-negative. 
We can also write an expression depending only on $r_0$ and $a$
\begin{equation}
\keff
=
\frac{\sqrt{3\,\Delta_{0}}}{r_{0}\,\sqrt{a^2+3r_{0}^2}}.
\label{eq:kappa-compact-2}
\end{equation}
The Rindler form immediately yields the instability exponent in boundary time $t$.  Consider a null curve in $(t,\rho)$ and set  ${\rm d}s^2_{(2)}=0$ in Eq. \eqref{eq:rindler} to get
\begin{equation}
\left(\frac{{\rm d}\rho}{dt}\right)^2\approx \keff^2\rho^2.
\end{equation}
Thus
\begin{equation}
\frac{{\rm d}\rho}{dt}\approx \pm \keff\,\rho
\quad\Rightarrow\quad
\rho(t)=\rho_0\,e^{\pm \keff t}.
\end{equation}
Hence the Lyapunov exponent is
\begin{equation}
\lambda=\keff=2\pi \Teff,
\label{eq:lambda-kappa}
\end{equation}
where $\keff$ is given by Eq. (\ref{eq:kappa-compact-01}) or equivalently by Eq. (\ref{eq:kappa-compact-2}). The same relation holds for $\lambda_{\rm p}$ on the equatorial photon ring, where $\dd t/\dd\tau_{\rm M}$ is constant. The Rindler wedge in Mino time gives
\be 
\lambda_{\rm p}=\kappa_{\rm eff}^{({\rm M})}=2\pi T_{\rm eff}^{({\rm M})}=\sqrt{3} r_0~,
\ee
where $\kappa_{\rm eff}^{({\rm M})}$ and $T_{\rm eff}^{({\rm M})}$ are  the corresponding quantities in Mino time along the orbit. 
Therefore, an analog to the classical maximal chaos bound is realized at the photon ring for Kerr black holes. There exists a local Rindler structure in the co-rotating reduced geometry near the photon ring, with instability exponent $\lambda$ and an effective temperature $T_{\rm eff}$, yielding the formal relation Eq. \eqref{eq:lambda-kappa}, analogous to chaos bound saturation on the horizon \cite{Maldacena:2015waa}.

\setcounter{equation}{0}
\setcounter{figure}{0}
\setcounter{section}{0}
\setcounter{table}{0}
\renewcommand{\theequation}{C\arabic{equation}}
\renewcommand{\thefigure}{C\arabic{figure}}
\renewcommand{\thetable}{C\arabic{table}}
\vspace{-0.5cm}
\section{C. The radial and polar potentials and  the Mino time}\label{supp:1}
\noindent
In this section we collect the ingredients entering  Eq. (\ref{eq::Lyapunov_Exponent}), restoring the dependence on the energy. 
We define the radial and angular potentials as
\begin{align}
    \mathcal{R}(r)&=\left[E (r^2+a^2)- a \ell \right]^2- k \Delta\nonumber\\
    \Theta(\theta)&= k - \left(\ell \csc\theta - a E \sin \theta\right)^2,
\end{align}
where $E, \ell, k$ are the three conserved quantities (energy, angular momentum and separation Carter-related constant). 
We can link these constants to the four-momentum of the test particle $p^{\mu}$ which follows a geodesic, obtaining
\begin{equation}
    E= -p_t,~~ \ell = p_{\mu}\partial_{\phi}^{\mu}=p_{\phi}, ~~ k = p_{\theta}^2 + (p_{\phi}\csc \theta + p_t a \sin \theta)^2.
\end{equation}
We highlight that in the text we have normalized everything to $E=1$.
The equations for the geodesic trajectories are
\begin{align}
\Sigma \frac{\dd r}{\dd\sigma} &= \pm_ r \sqrt{\mathcal{R}(r)}, \\ \label{Theta_dynamic}
\Sigma \frac{\dd \theta}{\dd \sigma} &= \pm_\theta \sqrt{\Theta(\theta)}, \\
\Sigma \frac{\dd \phi}{\dd \sigma} &= \frac{a}{\Delta}\left[E (r^2 + a^2) - a\ell \right]
+ \ell \csc^2\theta - aE, \\
\Sigma \frac{\dd t}{\dd \sigma} &= \frac{(r^2 + a^2)}{\Delta}
\left[E (r^2 + a^2) - a\ell \right]
+ a(\ell - aE \sin^2\theta),
\end{align}
where $\sigma$ is the affine parameter for massless particles.
From the above equations, we can identify the Mino time, defined as
\begin{align}
    \dd\tau_{\rm M}= \frac{\dd\sigma}{\Sigma},
\end{align}
which is related to the time in 
Boyer--Lindquist coordinates as \cite{Mino:2003yg}
\be
\frac{{\rm d}t}{{\rm d}\tau_{\rm M}}=\frac{r^2+a^2}{\Delta}\left[E(r^2+a^2)-a\ell\right]+ a(\ell - aE \sin^2\theta).
\ee

For spherical null orbits (imposing $\mathcal{R}(r)= \mathcal{R}'(r)=0$) it is straightforward to obtain
\begin{equation}
    \ell(r)= \frac{E\left[M(r^2-a^2)-r\Delta\right]}{a(r-M)},
\end{equation}
and
\begin{equation}
    k(r)= \frac{\left[E(r^2+a^2)-a\ell\right]^2}{\Delta},
\end{equation}
for
\begin{align}
r_- \le r \le r_+,\quad
\theta_- \le \theta \le \theta_+,\quad
0 \le \phi < 2\pi.
\end{align}
From Eq. (\ref{Theta_dynamic}) we can obtain the value of the average Mino time by integrating the angular potential between the two inversion points of the orbit \cite{Kapec:2019hro}
\begin{eqnarray}\nonumber
    \tau_{\rm M}(r) &=& \int_{\theta_{-}}^{\theta_+}
\frac{{\rm d}\theta}{\sqrt{\Theta(\theta)}} = \frac{2}{\sqrt{-u_{-}(r)a^2E^2}}K\left(\frac{u_{+}(r)}{u_{-}(r)}\right),
\end{eqnarray}
where $K$ is the elliptic integral of the first kind.
Note that the spinless limit, $\dd\tau_{\rm M}/\dd t=1/27 M^2=\lambda^2$ where $\lambda=1/(3\sqrt{3} M)$ is the Lyapunov coefficient of the photon ring at $r=3M$ of the Schwarzschild black hole, is such that $\overline{\lambda}_{\rm p}=1/\lambda$ in Mino time. 

\setcounter{equation}{0}
\setcounter{figure}{0}
\setcounter{section}{0}
\setcounter{table}{0}
\renewcommand{\theequation}{D\arabic{equation}}
\renewcommand{\thefigure}{D\arabic{figure}}
\renewcommand{\thetable}{D\arabic{table}}
 \vspace{-0.5cm}
\section{D. Impact of the unknown fifth-order post-Newtonian correction}\label{supp:2}
\noindent
To estimate the impact of unknown higher-order PN corrections in our determination of $j_*\nu$ (the location of the $\overline{\lambda}_{\rm p}$ maximum) we parametrize Eq.~(\ref{eq:Mj}) as
\begin{equation}
\frac{M(j)}{M_{\rm tot}}=1-\frac{\nu}{2j^2} \left[1+\sum_{n=1}^\infty c_n \, \frac{n}{2} \Big( \frac{2}{j}\Big)^{2n} \right] .
\end{equation}
This parametrization has been chosen to ensure that the PN coefficients $c_n$ remain order-one for all known corrections up to $n=4$. Indeed, we find that $(c_1, \dots , c_4)$ are $(1.15,0.62,0.54,0.60)$ for $q=1$, and $(1.14,0.62,0.58,0.71)$ for $q=6$.

Next, we calculate the value of $c_5$ required to shift the location of the maximum by a fixed fractional amount compared to the fit of Eq.\,\ref{fit}. For $q=1$, we find that $c_5 = (0.30,0.65,1.13)$ shift the values of $j_*\nu$ by $(1\%,2\%,3\%)$, respectively. For $q=6$, the same shifts are achieved for $c_5 = (0.52,1.18,1.98)$. This shows that the approximation of neglecting the fifth PN order introduces an uncertainty in our determination of $J/M_{\rm tot}^2$ which could be as large as a few percent.

\setcounter{equation}{0}
\setcounter{figure}{0}
\setcounter{section}{0}
\setcounter{table}{0}
\renewcommand{\theequation}{E\arabic{equation}}
\renewcommand{\thefigure}{E\arabic{figure}}
\renewcommand{\thetable}{E\arabic{table}}
\section{E. SXS catalog comparison}\label{app:sxs}
\noindent
We compare our prediction against direct numerical relativity (NR) data from the SXS Collaboration's third binary black hole catalog~\cite{Scheel:2025jct}. Starting from 3,756 simulations, we retain only those satisfying all three of the following selection criteria: \textbf{(a)} $|\chi_1|, |\chi_2| < 10^{-4}$, so that both initial black holes are effectively non-spinning; \textbf{(b)} eccentricity $e < 10^{-3}$ at the reference time, ensuring quasi-circular initial conditions; \textbf{(c)} non-deprecated runs, to exclude simulations with known numerical pathologies. Restricting to $q \lesssim 10$ yields 60 simulations, listed in Table~\ref{fig:App} together with the corresponding predictions from maximizing $\overline{\lambda}_{\rm p}$.
 The relative deviation is defined as
\begin{equation}
|\delta| \equiv \left|\frac{(J_f/M_{\rm tot}^2)_{\rm SXS} - \nu j_*}{\nu j_*}\right|,
\end{equation}
where $(J_f/M_{\rm tot}^2)_{\rm SXS}$ is the dimensionless remnant spin reported in the catalog and $\nu j_*$ is the value of the angular momentum at the maximum of $\overline{\lambda}_{\rm p}$, properly rescaled to $M_{\rm tot}$. As can be seen from the table, $|\delta|$ is consistently of order $2$--$3\%$ throughout the range $1 \le q \lesssim 10$, with the exception of a small set of simulations near $q \approx 2$ where the agreement improves to below $0.5\%$.
\newpage
\newpage
\begin{table*}[t!]
\centering
\caption{Comparison between the SXS catalog remnant spin $(J_f/M_{\rm tot}^2)_{\rm SXS}$ and the prediction $\nu j_*$ obtained by maximizing $\overline{\lambda}_{\rm p}$, for the $60$ non-spinning, quasi-circular, non-deprecated simulations with $q\lesssim 10$ selected from Ref.~\cite{Scheel:2025jct} (see text for selection criteria). The relative deviation $|\delta|$ is defined in the text. SXS IDs are abbreviated; the full identifier is \texttt{SXS:BBH:}$\langle$ID$\rangle$.}
\label{fig:App}
\scriptsize
\setlength{\tabcolsep}{4pt}
\begin{tabular}{ccccc @{\hspace{18pt}} ccccc}
\hline\hline
SXS ID & $q$ & $(J_f/M_{\rm tot}^2)_{\rm SXS}$ & $\nu j_*$ & $|\delta|\,\%$ &
SXS ID & $q$ & $(J_f/M_{\rm tot}^2)_{\rm SXS}$ & $\nu j_*$ & $|\delta|\,\%$ \\
\hline
0389 & 1.00 & 0.621595 & 0.637322 & 2.47 & 2485 & 4.00 & 0.450921 & 0.440118 & 2.45 \\
2377 & 1.00 & 0.621618 & 0.637322 & 2.46 & 3631 & 4.00 & 0.450891 & 0.440118 & 2.45 \\
2378 & 1.00 & 0.621616 & 0.637322 & 2.46 & 2499 & 4.00 & 0.451050 & 0.440118 & 2.48 \\
3624 & 1.00 & 0.621579 & 0.637322 & 2.47 & 1220 & 4.00 & 0.451089 & 0.440118 & 2.49 \\
4434 & 1.00 & 0.621611 & 0.637322 & 2.47 & 2484 & 4.50 & 0.425382 & 0.413996 & 2.75 \\
2496 & 1.00 & 0.621610 & 0.637322 & 2.47 & 3144 & 4.50 & 0.425363 & 0.413996 & 2.75 \\
1132 & 1.00 & 0.621607 & 0.637322 & 2.47 & 2374 & 5.00 & 0.401999 & 0.390755 & 2.88 \\
3864 & 1.00 & 0.621632 & 0.637322 & 2.46 & 2487 & 5.00 & 0.401998 & 0.390755 & 2.88 \\
1154 & 1.00 & 0.621605 & 0.637322 & 2.47 & 3619 & 5.00 & 0.402060 & 0.390755 & 2.89 \\
1155 & 1.00 & 0.621605 & 0.637322 & 2.47 & 2486 & 5.50 & 0.380942 & 0.370024 & 2.95 \\
1153 & 1.00 & 0.621606 & 0.637322 & 2.47 & 2164 & 6.00 & 0.361740 & 0.351429 & 2.93 \\
3617 & 1.00 & 0.621617 & 0.637322 & 2.46 & 2489 & 6.00 & 0.361739 & 0.351429 & 2.93 \\
2375 & 1.00 & 0.621616 & 0.637322 & 2.46 & 3630 & 6.00 & 0.361752 & 0.351429 & 2.94 \\
3634 & 1.00 & 0.621619 & 0.637322 & 2.46 & 2488 & 6.50 & 0.344264 & 0.334699 & 2.86 \\
2376 & 1.00 & 0.621616 & 0.637322 & 2.46 & 0192 & 6.58 & 0.341615 & 0.332174 & 2.84 \\
2325 & 1.00 & 0.621616 & 0.637322 & 2.46 & 2491 & 7.00 & 0.328335 & 0.319554 & 2.75 \\
2326 & 1.00 & 0.621619 & 0.637322 & 2.46 & 0188 & 7.19 & 0.322721 & 0.314172 & 2.72 \\
0198 & 1.20 & 0.618371 & 0.633042 & 2.32 & 2490 & 7.50 & 0.313570 & 0.305793 & 2.54 \\
2331 & 1.50 & 0.605941 & 0.616488 & 1.71 & 0195 & 7.76 & 0.306575 & 0.299126 & 2.49 \\
3984 & 1.50 & 0.606013 & 0.616488 & 1.70 & 2493 & 8.00 & 0.300277 & 0.293259 & 2.39 \\
0593 & 1.50 & 0.605969 & 0.616488 & 1.71 & 2707 & 8.00 & 0.300306 & 0.293259 & 2.40 \\
2497 & 2.00 & 0.576020 & 0.578756 & 0.47 & 3622 & 8.00 & 0.300415 & 0.293259 & 2.44 \\
1166 & 2.00 & 0.576082 & 0.578756 & 0.46 & 0186 & 8.27 & 0.293632 & 0.286931 & 2.34 \\
1167 & 2.00 & 0.576026 & 0.578756 & 0.47 & 2492 & 8.50 & 0.287965 & 0.281767 & 2.20 \\
0201 & 2.32 & 0.555024 & 0.553363 & 0.30 & 0199 & 8.73 & 0.282724 & 0.276799 & 2.14 \\
1178 & 3.00 & 0.508425 & 0.502630 & 1.15 & 2495 & 9.00 & 0.276604 & 0.271202 & 1.99 \\
1179 & 3.00 & 0.507720 & 0.502630 & 1.01 & 0189 & 9.17 & 0.273059 & 0.267802 & 1.96 \\
2498 & 3.00 & 0.509977 & 0.502630 & 1.46 & 1108 & 9.20 & 0.272235 & 0.267212 & 1.88 \\
2265 & 3.00 & 0.509816 & 0.502630 & 1.43 & 2494 & 9.50 & 0.266002 & 0.261460 & 1.74 \\
2483 & 3.50 & 0.479330 & 0.469554 & 2.08 & 0185 & 9.99 & 0.256561 & 0.252645 & 1.55 \\
\hline\hline
\end{tabular}
\end{table*}
\end{document}